\documentclass[preprint,showpacs,preprintnumbers,amsmath,amssymb,aps]{revtex4}
\usepackage{epsfig}
\usepackage{psfrag}
\usepackage{graphicx}
\usepackage{color,graphics}

\begin{document}

\title{Influence of flavor oscillations on neutrino beam instabilities}

\author{J.T. Mendon\c{c}a}
\email{titomend@ist.utl.pt}
\affiliation{IPFN, Instituto Superior T\'ecnico, 1049-001 Lisboa, Portugal}

\author{F. Haas}
\affiliation{Instituto de F\'isica, Universidade Federal do Rio Grande do Sul, CEP 91501-970, Porto Alegre RS, Brasil}

\author{A. Bret}
\affiliation{ETSI Industriales, Universidad de Castilla-La Mancha, 13071 Ciudad Real, Spain and
Instituto de Investigaciones Energeticas y Aplicaciones Industriales,
Campus Universitario de Ciudad Real, 13071 Ciudad Real, Spain}

\begin{abstract}
We consider the collective neutrino plasma interactions, and study the electron plasma instabilities produced by a nearly mono-energetic neutrino beam in a plasma. We describe the mutual influence of neutrino flavor oscillations and electron plasma waves. We show that the neutrino flavor oscillations are not only perturbed by electron plasmas waves, but also contribute to the dispersion relation and the growth rates of neutrino beam instabilities.

\end{abstract}


\maketitle

\section{Introduction}

Neutrino interactions with plasma are very important to understand supernova explosions and many other astrophysical phenomena \cite{raffelt}. Two types of effects arise from such interactions. First, they modify the neutrino flavor oscillations \cite{duan}, and introduce a resonant coupling between different flavor states, known as the MSW (Mikheyev-Smirnov-Wolfenstein) effect \cite{bethe,wolf,mikhey}. Second, they create an induced neutrino charge \cite{semikoz,serbeto}, which can lead to collective plasma oscillations and significantly increase the collision cross sections. The energy transfer between a neutrino beam and plasma wave is mediated by the neutrino Landau damping \cite{prlneu}. The individual flavor processes are important to understand the solar neutrino deficit, while the collective plasma effects could play a major role in supernova explosions \cite{review}. It should be noticed that the core-collapse problem is still unsolved \cite{burrows}. A possible solution could eventually be given by these plasma effects.

In a recent work, we have proposed to built a bridge between these two kinds of phenomena, and have introduced  plasma physics methods in the discussion of neutrino flavor oscillations in matter. In particular, we have derived exact BGK (Bernstein-Greene-Kruskal) like  solutions for electron plasma density profiles compatible with given neutrino flavor parameters \cite{haas}, and have determined the modified neutrino flavor oscillations in the presence of plasma waves and turbulence \cite{mend}. Here, we take a further step in the same direction, by considering the mutual influence of flavor oscillations and plasma instabilities. We will show that the neutrino flavor parameters are not only perturbed by electron plasmas oscillations, but can also contribute to both the dispersion relation and the growth rates of neutrino beam instabilities in a plasma.

This paper is organized in the following way. In Section II, we consider the basic equations of our problem, by considering a simple fluid description where ions are assumed at rest and the neutrino flavor oscillations are taken into account. In Section III, we consider a plasma in steady state, and characterize the unperturbed solutions for both the electron plasma parameters and the  flavor polarization vector. In Section IV, we consider the perturbations induced by a plasma wave and establish the evolution equations for the perturbed quantities. From this perturbative analysis it becomes clear that plasma waves induce perturbation in the neutrino flavor parameters which, on the other hand, lead to new dispersive effects. In Section V, we consider the case of electron plasma waves excited by an incoherent neutrino beam, and determine the corresponding growth rates. In Section VI we generalize this analysis to the case of a coherent neutrino beam and show that the flavor oscillations can contribute to the dispersion relation of electron plasma waves. Finally, in Section VII, we state some conclusions.

\section{Fluid description}

We assume a simple fluid description for both the plasma electrons and the neutrino beam, with immobile ions. A unit system with $\hbar = c = 1$ will be used. The electrons are described by the non-relativistic fluid equations fluid equations
\begin{equation}
\frac{\partial n}{\partial t} + \nabla \cdot (n {\bf v}) = 0 \; , \quad
\frac{\partial {\bf p}}{\partial t} + {\bf v} \cdot  \nabla {\bf p} = {\bf F} + {\bf F}_\nu - \frac{\nabla P}{n} \, ,
\label{2.1} \end{equation}
where $n$ and ${\bf v}$ are the electron mean density and velocity,
${\bf p} = m {\bf v}$ and $P$ is the electron pressure.
We have also used the Lorentz force ${\bf F}$, and the neutrino force ${\bf F}_\nu$, defined by
\begin{equation}
{\bf F} = - e ({\bf E} + {\bf v} \times {\bf B}) \; , \quad {\bf F}_\nu = \sqrt{2} G_F ({\bf E}_\nu + {\bf v} \times {\bf B}_\nu) \, ,
\label{2.2} \end{equation}
where ${\bf E}$ and ${\bf B}$ are the electric and magnetic fields, $e$ is the electron charge, $G_F$ the Fermi constant of weak interactions, and ${\bf E}_\nu$ and ${\bf B}_\nu$ are effective fields induced by the weak interactions, determined by
\begin{equation}
{\bf E}_\nu = - \nabla N_e - \frac{\partial {\bf J}_e}{\partial t} \; , \quad {\bf B}_\nu = \nabla \times {\bf J}_e \, .
\label{2.3} \end{equation}
The electron-neutrino density $N_e$, and current ${\bf J}_e = N_e {\bf v}_e $, are coupled to the muon neutrino density $N_\mu$ and current ${\bf J}_\mu =  N_\mu {\bf v}_\mu$, as shown by the continuity equations
\begin{equation}
\frac{\partial N_e}{\partial t} + \nabla \cdot  {\bf J}_e =  \frac{N_0}{2} \Omega_0 P_2 \; , \quad
\frac{\partial N_\mu}{\partial t} + \nabla \cdot  {\bf J}_\mu = - \frac{N_0}{2} \Omega_0 P_2 \, ,
\label{2.4} \end{equation}
where the quantity $P_2$ pertains to neutrino coherence, as clarified below. Here, for simplicity, we restrict our analysis to the familiar two-flavor model \cite{raffelt}, but extension to the three neutrino flavor states would not be difficult. In these two coupled equations we have also used the constants
\begin{equation}
\Omega_0 = \omega_0 \sin 2 \theta_0 \; , \quad N_0 = N_e + N_\mu \, .
\label{2.5} \end{equation}
Strictly speaking, Eq. (\ref{2.4}) only apply to a coherent neutrino beam with a defined energy $\mathcal{E}_0$, and cannot be used in arbitrary situations. For this reason, we will only restrict our discussion to the neutrino beam interaction with a plasma, where the frequency $\omega_0$  can be unequivocally defined as $\omega_0 = \Delta m^2 / 2 \mathcal{E}_0$, and $\Delta m^2$ is the neutrino square mass difference. On the other hand, the quantum coherence factor $P_2$ satisfies the relations
\begin{equation}
\frac{d P_1}{d t} = - \Omega  P_2
 \; , \quad \frac{d P_2}{d t} = \Omega P_1 - \frac{\Omega_0}{N_0} (N_e - N_\mu)
  \, ,
\label{2.6}
\end{equation}
where we have defined
\begin{equation}
\Omega  = \omega_0 ( \cos 2 \theta_0 - \zeta  ) \; , \quad \zeta  = \sqrt{2} G_F \frac{n}{\omega_0} \, ,
\label{2.7} \end{equation}
The meaning of the total time derivative in Eq. (\ref{2.6}) will be clarified in Section III. To complete the description of the neutrino populations, we should consider the neutrino momentum equations
\begin{equation}
\frac{\partial {\bf p}_e}{\partial t} + {\bf v}_e \cdot  \nabla {\bf p}_e = \sqrt{2} G_F ({\bf E}_e + {\bf v}_e \times {\bf B}_e)
\; , \quad \frac{\partial {\bf p}_\mu}{\partial t} + {\bf v}_\mu \cdot  \nabla {\bf p}_\mu = 0 \, ,
\label{2.8} \end{equation}
with ${\bf p}_e = {\bf v}_e \mathcal{E}_e$ and ${\bf p}_\mu = {\bf v}_\mu \mathcal{E}_\mu$. The quantities ${\bf E}_e$ and ${\bf B}_e$ appearing in these equations are effective fields \cite{serbeto} defined as
\begin{equation}
{\bf E}_e = - \nabla n - \frac{\partial}{\partial t} (n {\bf v})  \; , \quad {\bf B}_e = \nabla \times (n {\bf v}) \, .
\label{2.8b} \end{equation}
The quantity $\Omega_0$ in Eq. (\ref{2.5}) depends on the energy of the neutrino beam, and is well defined for a (nearly mono-energetic) beam, such that $\mathcal{E}_e = \mathcal{E}_\mu = \mathcal{E}_0$. The momentum equations in (\ref{2.8}) show that, in the presence of plasma perturbations ${\bf E} \neq 0$ and ${\bf B} \neq 0$, the quantities ${\bf v}_e$ and ${\bf v}_\mu$ are not necessarily identical.

In conclusion, we have a rather detailed model for the neutrino-plasma coupling. The electron variables $n$ and ${\bf v}$ are determined in a self-consistent way together with the fields ${\bf E}, {\bf B}$ through the Maxwell equations with self-consistent charge and current density, with a coupling produced by the neutrino force ${\bf F}_\nu$ in the electron momentum equation (\ref{2.1}). The neutrino force depends on the effective neutrino fields ${\bf E}_\nu, {\bf B}_\nu$, which in turn are specified by the electron neutrino variables $N_e, {\bf v}_e$. However, in the present two-flavor model, the electron neutrino variables oscillate due to the quantum coherence $P_2$, coupling with the muon neutrino quantities $N_\mu, {\bf v}_\mu$. Finally, the neutrino oscillations are influenced by the plasma in two ways: due to the coupling with $n$ in $\zeta$ in Eq. (\ref{2.7}) and due to the role of the electromagnetic field in the electron neutrino momentum equation (\ref{2.8}).
In the following, illustrative examples of applications are provided.

\section{Equilibrium state}

For simplicity, let us consider a non-magnetized plasma,  although e.g. regarding a Supernovae setting future works accounting for a non-zero ambient magnetic field could be quite relevant. In equilibrium, we have $n = n_0$,  ${\bf v} = 0$ and ${\bf E} = {\bf B} = 0$. This implies that, for a mono-energetic neutrino beam, the velocity of both flavors are identical, and we can use ${\bf v}_e = {\bf v}_\mu= {\bf v}_0$. In this case, Eq. (\ref{2.4}) will reduce to the following relation
\begin{equation}
\frac{d P_3}{d t} \equiv \left( \frac{\partial}{\partial t} + {\bf v}_0 \cdot \nabla \right) P_3 = \Omega_0 P_2 \, ,
\label{3.1} \end{equation}
with $P_3 = (N_e - N_\mu) / N_0$. In this case, the coupled equations (\ref{2.6}) and (\ref{3.1}) can describe the evolution of a three-dimensional flavor polarization vector ${\bf P} \equiv (P_1, P_2, P_3)$, where the total time derivatives are defined without ambiguity. They can be rewritten as
\begin{equation}
\frac{d P_1}{d t} = - \bar \Omega P_2 \,, \quad  \frac{d P_2}{d t} =  \bar \Omega P_1 - \Omega_0 P_3 \,, \quad \frac{d P_3}{d t} = \Omega_0 P_2 \,,
\label{3.2} \end{equation}
with $\bar \Omega = \omega_0 \cos 2 \theta_0 - \sqrt{2} G_F n_0$. This implies that
\begin{equation}
\frac{d^2 P_2}{d t^2} = - \bar \omega^2 P_2 \; , \quad \bar \omega^2 = \bar \Omega^2 + \Omega_0^2 \, .
\label{3.3} \end{equation}
Introducing a new angle $\bar \theta$, we can also write
\begin{equation}
\bar \omega = \frac{\Omega_0}{\sin 2 \bar \theta} \; , \quad \tan 2 \bar \theta = \frac{\Omega_0}{\bar \Omega} \, .
\label{3.4} \end{equation}
Equation (\ref{3.3}) can be solved as
\begin{equation}
P_2 (t) = P_{20}(t) = A \exp (- i \bar \omega t) + B \exp (i \bar \omega t) \, .
\label{3.5} \end{equation}
It is convenient to define the constants of integration as
$A = - B = (\beta/ 2 i) \sin 2 \bar \theta$, in terms of a free parameter $\beta$ related to quantum coherence, leading to
\begin{equation}
P_{20} (t) = - \beta\sin 2 \bar \theta \sin  \bar \omega t \, .
\label{3.5b} \end{equation}
Accordingly, without loss of generality a phase constant was chosen so that $P_{20}(0) = 0$.
For completeness, we show the results for the remaining polarization vector components, compatible with the normalization condition $|{\bf P}| = 1$,
\begin{equation}
\label{full}
P_{10}(t) = \sin\phi - \beta \sin 2\bar\theta \cos 2\bar\theta (\cos \bar\omega t - 1) \,, \quad P_{30}(t) = \cos\phi + \beta \sin^{2} 2\bar\theta (\cos \bar\omega t -1 ) \,,
\end{equation}
where the angle $\phi$ satisfy
\begin{equation}
\sin(2\bar\theta - \phi) = \beta \sin 2 \bar\theta \cos 2 \bar\theta \,.
\end{equation}
For instance, in the case of absence of quantum coherence ($\beta = 0$) one has $\phi = 2\bar\theta$. We can explicitly see that $\beta$ is a measure of the amplitude of the flavor oscillations, for if $\beta = 0$ there are no oscillations at all.

The corresponding solution for the equilibrium electron-neutrino density is
\begin{equation}
N_{e0} (t) = N_0 \left[1 - \frac{\beta}{2} \sin^2 2 \bar \theta \left(1 - \cos \bar \omega t \right) \right] \, ,
\label{3.6} \end{equation}
and $N_{\mu 0} (t) = N_0 - N_{e0} (t)$.
These results are strictly valid for a mono-energetic beam $\Delta \mathcal{E}_0 \ll \mathcal{E}_0$. A finite value of $\Delta \mathcal{E}_0$ would introduce a temporal (and/or spatial) decay of these oscillations \cite{raffelt}.

Nothing that Eq. (\ref{3.2}) contain total derivatives
(in a Lagrangian variables sense), we can also consider another class of initial conditions,  $N_e ({\bf r} = 0) = N_0$ and $N_\mu ({\bf r} = 0) = 0$. The resulting solutions would then be expressed in terms of spatial coordinates as
\begin{equation}
P_2 ({\bf r}) = A \exp (- i {\bf k}_0 \cdot  {\bf r}) + B \exp (i {\bf k}_0 \cdot  {\bf r}) \, ,
\label{3.7} \end{equation}
where ${\bf k}_0 = (\bar \omega / v_0^2) {\bf v}_0$. In the following, we will focus on equilibrium solutions of the type (\ref{3.5}), although a similar analysis could be done for solutions of the type (\ref{3.7}).

\section{Perturbative analysis}

In the previous Section there was no electromagnetic field at all.
We now consider the possible excitation of electron plasma waves by a mono-energetic neutrino beam, as described by the perturbed densities $\tilde n = n - n_0$, $\tilde N_e = N_e - N_{e0}$, and $\tilde N_\mu = N_\mu - N_{\mu 0}$. We restrict our discussion to electrostatic waves, with ${\bf B} = 0$ and ${\bf E}$ determined by Poisson's equation
\begin{equation}
\nabla \cdot {\bf E} = - \frac{e}{\epsilon_0} \tilde n \, .
\label{4.1} \end{equation}
Linearizing the electron fluid Eqs. (\ref{2.1}) and (\ref{2.2}), with ${\bf B}_\nu \simeq 0$, we get
\begin{equation}
\frac{\partial \tilde n}{\partial t} + n_0 \nabla\cdot{\bf v} = 0 \; , \quad
\frac{\partial {\bf p}}{\partial t} = - e {\bf E} - \sqrt{2} G_F \nabla \tilde N_e  - \frac{\nabla \tilde P}{n_0} \, .
\label{4.2} \end{equation}
In addition, the displacement current contribution from (\ref{2.3}) has been disregarded due to the non-relativistic assumption.
Taking the time derivative of the continuity equation and using (\ref{4.1})
we obtain
\begin{equation}
\left( \frac{\partial^2}{\partial t^2} + \omega_p^2 - v_{th}^2 \nabla^2 \right) \tilde n  = \frac{\sqrt{2} n_0}{m} G_F \nabla^2 \tilde N_e \, ,
\label{4.3} \end{equation}
where we have used the electron plasma frequency $\omega_p = \sqrt{e^2 n_0 / \epsilon_0 m}$, and the electron thermal velocity  $v_{th} = \sqrt{3 T / m}$, where $T$ is the electron temperature. In this expression, we can easily recognize the usual wave equation for electron density perturbations, with an additional term associated with the electron-neutrino oscillations.

In order to determine the quantity $\tilde N_e$, we linearize Eqs. (\ref{2.4}) - (\ref{2.7}), assuming that the unperturbed neutrino flavor solutions $N_{e 0}$, $N_{\mu 0}$ and $P_{2 0}$ are given by the (homogeneous in space) solutions of Section III. This leads to
\begin{equation}
\frac{d \tilde N_e}{d t} + \nabla \cdot  ( N_{e 0} \tilde {\bf v}_e) =  \frac{N_0}{2} \Omega_0 \tilde P_2 \; , \quad
\frac{d \tilde  N_\mu}{d t}  + \nabla \cdot (N_{\mu 0} \tilde {\bf v}_\mu)  = - \frac{N_0}{2} \Omega_0 \tilde P_2 \, ,
\label{4.4} \end{equation}
complemented by the neutrino momentum equations
\begin{equation}
\frac{d \tilde {\bf p}_e}{d t} = \sqrt{2} G_F {\bf E}_e \; , \quad
\frac{d \tilde {\bf p}_\mu}{d t}  = 0 \, .
\label{4.4b} \end{equation}
In these expressions we have used the total time derivative $d / d t \equiv (\partial / \partial t + {\bf v}_0 \cdot \nabla)$. We can immediately recognize that the muon-neutrinos are not accelerated by the electron plasma wave, because they are not directly coupled to the electron perturbations. This is strictly valid only for a neutral electron-proton plasma. We should also note that $\tilde {\bf v}_e =  \tilde {\bf p}_e/ \mathcal{E}_0$,where $\mathcal{E}_0$ is the unperturbed neutrino beam energy. From here, we can derive an evolution equation for the perturbed electron-neutrino density, of the form
\begin{equation}
\frac{d^2 \tilde N_e}{d t^2} + \frac{d N_{e 0}}{d t} \nabla \cdot \tilde {\bf v}_e - \alpha_p^2 \nabla^2 \tilde n =  \frac{N_0}{2} \Omega_0 \frac{d \tilde P_2}{d t}  \, ,
\label{4.5} \end{equation}
with
\begin{equation}
\alpha_p^2 = \sqrt{2} G_F  \frac{N_{e0}}{\mathcal{E}_0}  \, .
\label{4.6} \end{equation}
It can be seen that the evolution of the perturbed density $\tilde N_e$ is coupled to $\tilde n$, and also depends on the perturbed coherence $\tilde P_2$. An equation for this quantity can be derived from
Eq. (\ref{2.6}), and can be written as
\begin{equation}
\frac{d^2 \tilde P_2}{d t^2} + \bar \Omega^2\tilde P_2 =  \tilde \Omega \left( \frac{d P_{10}}{d t} - \bar \Omega P_{20} \right) - \frac{\Omega_0}{N_0} \frac{d}{d t} (\tilde N_e - \tilde N_\mu) + P_{10} \frac{d \tilde \Omega}{d t} \,,
\label{4.6b} \end{equation}
where it was defined $\tilde \Omega = \Omega - \bar \Omega$.
Using the above equations (\ref{4.4}) and (\ref{4.4b}), and taking the definition of $\bar \omega$ into account, we can then transform this into
\begin{equation}
\frac{d^2 \tilde P_2}{d t^2} + \bar \omega^2\tilde P_2 = - \sqrt{2} G_F \left[ \left( \frac{d P_{10}}{d t} - \bar \Omega P_{2 0} \right)  \tilde n + P_{10} \frac{d \tilde n}{d t} \right] + \frac{\Omega_0}{N_0} N_{e0}  \nabla \cdot \tilde {\bf v}_e   \, ,
\label{4.7} \end{equation}
where the perturbed neutrino velocity is determined by the equation of motion $d \tilde {\bf v}_e/dt = (1 / \mathcal{E}_0) \sqrt{2} G_F {\bf E}_e$. We now have all the equations for the perturbed quantities, which will be solved in the next two Sections.

\section{Incoherent neutrino beam}

Let us first consider the simple case of $P_{20} = 0$ and $\beta = 0$, which corresponds to the absence of quantum coherence.
In this case, we have two coupled equations for the variables $\tilde N_e$ and $\tilde n$.
On the other hand, no flavor oscillations will occur, and $d N_{e0} / d t = 0$. We therefore take $N_{e0} = N_0 = const$. For perturbations evolving as $\exp (i {\bf k} \cdot {\bf r} - i \omega t)$, we can reduce
Eqs. (\ref{4.3}) and (\ref{4.5}) to
\begin{equation}
\left( \omega^2 - \omega_p^2 - v_{th}^2 k^2 \right) \tilde n = \sqrt{2} G_F (n_0 / m) k^2 \tilde N_e \, ,
\label{5.1} \end{equation}
and
\begin{equation}
\left( \omega - {\bf k} \cdot {\bf v}_0 \right)^2  \tilde N_e =  \alpha_p^2 k^2 \tilde n \, .
\label{5.1b} \end{equation}
From here, we can derive the dispersion relation for plasma waves in the presence of an incoherent neutrino beam, as
\begin{equation}
\omega^2 - \omega_p^2 - v_{th}^2 k^2  =   \sqrt{2} G_F \frac{n_0}{m} \frac{k^4 \alpha_{p}^2}{(\omega - {\bf k} \cdot {\bf v}_0)^2} \, .
\label{5.1c} \end{equation}
This can also be written in the standard form, by introducing the plasma dielectric function $\epsilon (\omega, {\bf k})$, as
\begin{equation}
\epsilon (\omega, {\bf k}) \equiv 1 + \chi_e (\omega, {\bf k}) + \chi_\nu (\omega, {\bf k} ) = 0 \, ,
\label{5.2} \end{equation}
where the electron and neutrino susceptibilities are defined by the expressions
\begin{equation}
\chi_e (\omega, {\bf k}) = - \frac{1}{\omega^2} \left( \omega_p^2 + v_{th}^2 k^2 \right)
\; , \quad
\chi_\nu (\omega, {\bf k}) = -  \frac{{{\omega}^{4}_\nu}}{(\omega - {\bf k} \cdot {\bf v}_0)^2 \omega^2} \,, \quad  {\omega}^{4}_\nu = \frac{\sqrt{2} G_F n_0 k^4 \alpha_{p}^2}{m} \,.
\label{5.3} \end{equation}
This dispersion relation shows the possible occurrence of neutrino beam instabilities.
Maximum growth rates occur for the double resonance condition $\omega_r^2 = \omega_p^2 + v_{th}^2 k^2 = ({\bf k} \cdot {\bf v}_0)^2 $, where we have assumed $\omega = \omega_r + i \gamma$. The corresponding growth rate for the unstable solution is
\begin{equation}
\gamma = \frac{\sqrt{3}}{2} \, \omega_p \left(  \frac{N_{e0} n_0 G_F^2}{m \mathcal{E}_0 v_0^4} \right)^{1/3} \propto G_{F}^{2/3} \, .
\label{5.4} \end{equation}
valid for ${\bf k} \parallel {\bf v}_0$ and high frequency waves such hat $\omega_p^2 \gg k^2 v_{th}^2$. In the growth rate expression, the free parameters are $N_{e0}$ (the initial electron neutrino population), $\mathcal{E}_0$ (the defined coherent neutrino beam energy) and the neutrino beam velocity $v_0$ as well as the equilibrium electron density. Finally, it should be observed that Eq. (\ref{5.3}) is valid for $v_0 \neq 0$: it can easily be shown from Eq. (\ref{5.1b}) that there is no instability if $v_0 = 0$, as expected. Hence, the above growth rate estimate is always defined for the relevant scenarios. Also observe the growth rate is $\propto G_{F}^{2/3}$ which is the same estimate as in \cite{prlneu} for a weak neutrino beam.

Revealing insights can be obtained rewriting the dispersion relation in terms of a characteristic function $F(\omega)$ as
\begin{equation}
F(\omega) = \frac{\omega_{p}^2 + v_{th}^2 k^2}{\omega^2} +  \frac{{\omega_\nu}^4}{(\omega - {\bf k} \cdot {\bf v}_0)^2 \omega^2} \equiv 1 \,.
\label{cf}
\end{equation}
The dispersion relation is a fourth degree equation for $\omega$. Hence, to admit complex conjugate solutions and hence an unstable mode, the local minimum at $\omega = \omega^{*}$ (see figure 1) should satisfy $F(\omega^{*}) > 1$. To leading order in the neutrino coupling effects, we find
\begin{equation}
\omega^* = {\bf k} \cdot {\bf v}_0 - \frac{\omega_{\nu}^{4/3} ({\bf k}\cdot{\bf v}_0)^{1/3}}{\omega_{p}^{2/3}} \,,
\end{equation}
together with the instability condition
\begin{equation}
\label{ff}
F(\omega^*) = \frac{\omega_{p}^2 + v_{th}^2 k^2}{({\bf k} \cdot {\bf v}_0)^2} + \left(\frac{\omega_\nu \omega_p}{({\bf k}\cdot{\bf v}_0)^2}\right)^{4/3} > 1 \,,
\end{equation}
where ${\bf k}\cdot{\bf v}_0 > \omega_{\nu}^2$ was also assumed. Notice that the instability is enhanced by larger neutrino effects as apparent from the second term in Eq. (\ref{ff}). Moreover if the double resonance condition is fulfilled the instability condition is also immediately satisfied since the neutrino term is always positive (never stabilizing).

\begin{figure}[h]
 \includegraphics[width=6.0in,height=2.6in]{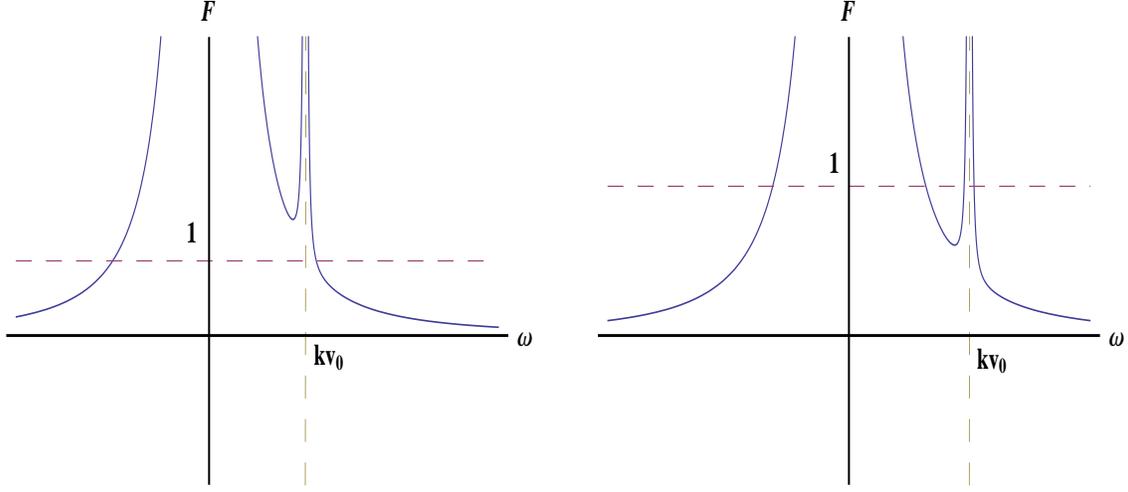}
  \caption{On the left: characteristic function from Eq. (\ref{cf}), in a generic unstable case such that $F(\omega^*) > 1$, where $\omega^*$ is the local minimum. There are only two real roots for the dispersion relation equation. On the right: the same but for a generic stable case such that $F(\omega^*) < 1$ and four real roots for the dispersion relation.}
 \label{x0}
\end{figure}

\section{Coherent neutrino beam}

Let us now consider the case of a coherent and nearly mono-energetic neutrino beam, where we have to retain the contributions of the coherence parameter $P_2$. To enhance this contribution, we suppose $\beta = 1, \phi = 0$ in the unperturbed solutions (\ref{3.5b})-(\ref{full}). We can then rewrite the coupled equations  (\ref{4.5}) and (\ref{4.7}) in the following form
\begin{equation}
\frac{d^2 \tilde N_e}{d t^2}
- \alpha_p^2 \nabla^2 \tilde n =  \frac{N_0}{2} \, \Omega_0 \frac{d \tilde P_2}{d t} \, ,
\label{6.1} \end{equation}
and
\begin{equation}
\frac{d^2 \tilde P_2}{d t^2} + \bar \omega^2\tilde P_2 + \left\{ a   \sin (\bar \omega t) + a' [ \cos (\bar \omega t) - 1 ] \right\}  \tilde n = 0\, ,
\label{6.2} \end{equation}
with the following auxiliary quantities
\begin{equation}
a =  2 \sqrt{2}  G_F \bar \Omega  \sin 2 \bar \theta \; , \quad a' = \frac{a}{2} \cos 2 \bar \theta \, .
\label{6.2b} \end{equation}
To simplify the analysis, we neglected the terms containing the perturbed electron and muon neutrino velocities.
We then assume perturbations of the form
\begin{equation}
(\tilde n, \tilde N_e, \tilde P_2) = \sum_l \left( \tilde n_l, \tilde N_{el}, \tilde P_{2l} \right) \exp \left( i {\bf k} \cdot {\bf r} - i \omega_l t \right)
\label{6.3} \end{equation}
with $\omega_l = \omega + l \bar \omega$, This allows us to establish the following relations
\begin{equation}
\left( \omega_l^2 - \omega_p^2 - v_{th}^2 k^2 \right) \tilde n_l = \sqrt{2} G_F \frac{n_0}{m} k^2 \tilde N_{e l}
\label{6.4a} \end{equation}
\begin{equation}
\left( \omega_l - {\bf k} \cdot {\bf v}_0 \right)^2 \tilde N_{el} = - \alpha_p^2 k^2 \tilde n_l + \frac{i}{2} (\omega_l - {\bf k} \cdot {\bf v}_0) N_0 \Omega_0 \tilde P_{2l}
\label{6.4b} \end{equation}
and
\begin{equation}
\left( \omega_l - {\bf k} \cdot {\bf v}_0 \right)^2 \tilde P_{2l} - \bar \omega^2 \tilde P_{2l} = \frac{i a}{2} \left( \tilde n_{l - 1} - \tilde n_{l + 1} \right) + \frac{a'}{2} \left( \tilde n_{l - 1} + \tilde n_{l + 1} - 2 \tilde n_l \right)
\label{6.4c} \end{equation}
From here, we obtain the recurrence relation
\begin{equation}
\omega_l^2 \epsilon (\omega_l, {\bf k}) \tilde n_l = - a A (\omega_l, {\bf k}) \left[ \left( \tilde n_{l - 1} - \tilde n_{l + 1} \right) - i \cos 2 \bar \theta \left( \tilde n_{l - 1} + \tilde n_{l + 1} - e \tilde n_l \right) \right] \,
\label{6.5} \end{equation}
where $ \epsilon (\omega_l, {\bf k})$ is given by Eq. (\ref{5.2}), with $\omega$ replaced by $\omega_l$, and the coupling function is defined by
\begin{equation}
A (\omega_l, {\bf k})= \frac{\sqrt{2} G_F}{4m} \frac{n_0 N_0 \Omega_0 k^2}{ ( \omega_l - {\bf k} \cdot {\bf v}_0) [( \omega_l - {\bf k} \cdot {\bf v}_0)^2 - \bar \omega^2]}
\label{6.6} \end{equation}
In order to study mode coupling contributions, we can use an approximation in Eq. (\ref{6.5}), by noting that the differences between $\tilde n_{l - 1}$ and $\tilde n_{l + 1}$ are very small,  if we assume that $l \bar \omega \ll \omega_p$. Using Eq. (\ref{6.4a}), we obtain by differentiation, $2 \omega_l  \tilde n_l \delta \omega_l + \omega_l^2 \delta \tilde n_l \simeq 0$. This leads to
\begin{equation}
\delta \tilde n_l \equiv \tilde n_{l-1} - \tilde n_{l+1} \simeq - 2 \frac{\delta \omega_l}{\omega_l} \tilde n_l
\label{6.6b} \end{equation}
Now, using $\delta \omega_l \equiv -  2 \bar \omega$, and noting that $\omega_l \simeq \omega_p$, we finally arrive at the simple estimate
\begin{equation}
 \tilde n_{l - 1} - \tilde n_{l + 1} \simeq  4 \frac{\bar \omega}{\omega_p} \tilde n_l
\label{6.7} \end{equation}
Taking $l = 0$, this leads to the following dispersion relation
\begin{equation}
\epsilon (\omega, {\bf k}) + \frac{4a}{\omega^2}  \frac{\bar \omega}{\omega_p} A (\omega, {\bf k}) = 0
\label{6.8} \end{equation}
This result shows that the previous dispersion relation (\ref{5.2}) for the neutrino beam interactions is corrected by an additional factor due to quantum correlations, which is of the order of $\bar \omega / \omega_p \ll 1$. Such corrections result from the coupling between quantum flavor oscillations and electron plasma waves, or in other words, between the quantum properties of neutrinos and their collective interactions with the plasma. Another interesting case, would be that of a near resonance between plasma and flavor oscillations, such that $\bar \omega \simeq \omega_p$, where stronger effects can be expected.

\section{Conclusions}

We have studied the influence the neutrino flavor oscillations on electron plasma waves created by neutrino beams, or in other words, the influence of the neutrino quantum properties on their collective behavior. The flavor oscillations result from the difference between the neutrino mass states and  their interaction states. On the other hand, the neutrino beam instabilities result form their weak coupling with the plasma. We have shown that the existence of flavor oscillations gives a new contribution to plasma dispersion relation, thus changing the frequency and growth rates of the beam instabilities.

In our model, we have introduced a number of simplifying assumptions. First, we have only retained the charged weak current and ignored the contributions from the neutral weak current. It is known that the electron neutrinos are coupled by the charged bosons $W^\pm$ to the electrons, and all neutrino flavors are coupled by the neutral boson $Z$ with both electrons and protons. For a plasma of electrons and protons in equilibrium, this weak coupling would give no net contribution to the neutrino-plasma interactions. In contrast, in the presence of a perturbation, this would lead to a correcting factor of order one to the terms proportional to $G_F$.

Furthermore, our work was only focused on electron plasma oscillations, and the protons (or ions) were assumed immobile. But we could also consider the excitation of ion acoustic waves, with frequencies of the order of the flavor oscillation frequency. These excitation could be driven by flavor oscillations. We should notice that the collective ion-neutrino coupling is mediated by the electrons, which can be assumed in Boltzmann equilibrium in the electrostatic and weak field potentials.

Turning now the the relevance of the present theory to core-collapse supernova, we can evaluate the instability growth rate. In SI units, Eq. (\ref{5.4}) reads,

\begin{equation}
\gamma = \frac{\sqrt{3}}{2} \, \omega_p \left(  \frac{N_{e0} n_0 G_F^2 c^2}{m \mathcal{E}_0 v_0^4} \right)^{1/3}.
\end{equation}

An electronic density $n_0=10^{35}$ m$^{-3}$ \cite{burrows} yields a plasma frequency $\omega_p=1.8\times 10^{19}$ s$^{-1}$. Considering also $N_{e0}=10^{35}$ m$^{-3}$, $\mathcal{E}_0 = 50$ MeV, $v_0 = c/10$ and with $G_F=1.45\times 10^{-62}$ J.m$^3$, one finds a growth rate,
\begin{equation}\label{eq:gr}
  \frac{\gamma}{\omega_p}=2.75\times 10^{-9}~\Rightarrow ~ \gamma^{-1} = 0.02 ~ \mathrm{ns},
\end{equation}
which is far shorter than the time scale ($\sim 1$ second) of the explosion. This instability definitely appears fast enough to alter neutrino mixing in core-collapse supernova.

The present results show that electron plasma waves excited by intense neutrino beams are intimately linked with the quantum processes associated with flavor oscillations. The dispersion relation and growth rates of these plasma instabilities are directly influenced by these flavor oscillations. It should however be noticed that we have also restricted our analysis to nearly mono-energetic neutrino beams. The general case of an arbitrary neutrino population can only be treated in the frame of a quantum statistical approach, which will be considered in a future publication.



\end{document}